\documentclass[9pt,onecolumn,twoside]{opticajnl}
\journal{opticajournal} % use for journal or Optica Open submissions

\setboolean{shortarticle}{false}
\usepackage{lineno}
\title{Interference of high-order perfect optical vortex beams}

\author[1,2,3,$\dagger$]{Bikash K. Das}
\author[1,2,3]{C. Granados}
\author[2,4]{M. Krüger}
\author[1,2,3,*]{M. F. Ciappina}
\affil[1]{Department of Physics, Guangdong Technion - Israel Institute of Technology, 241 Daxue Road, Shantou, Guangdong, China, 515063}
\affil[2]{Department of Physics, Technion -- Israel Institute of Technology, Haifa, 32000, Israel}
\affil[3]{Guangdong Provincial Key Laboratory of Materials and Technologies for Energy Conversion, Guangdong Technion - Israel Institute of Technology, 241 Daxue Road, Shantou, Guangdong, China, 515063}
\affil[4]{Solid State Institute, Technion -- Israel Institute of Technology, Haifa, 32000, Israel}

\affil[$\dagger$]{bikash.das@campus.technion.ac.il}
\affil[*]{marcelo.ciappina@gtiit.edu.cn}

\begin{abstract}
We investigate the interference of high-order perfect optical vortex (POV) beams with different topological charges. Through numerical simulations, we reveal a remarkable phenomenon: keeping the beam width, and beam radius fixed while changing the topological charge, the splitting of the composite POV beam into two distinct individual perfect vortices occurs exactly at the same inter-axial separation. The observed interference pattern exhibits pronounced sensitivity to factors such as axial separation, phase shift, beam radius, and topological charges of the constituent beams. Notably, our findings are contrasted with the interference of high-order Laguerre-Gauss (LG) beams, highlighting that the splitting of composite vortices into their individual components is more rapid in the case of LG beams. Our research provides significant insights into the distinct interference properties of high-order POV beams, presenting potential applications in the fields of optical manipulation and communication systems.

\end{abstract}

\setboolean{displaycopyright}{false} % Do not include copyright or licensing information in submission.

\begin{document}

\maketitle

\section{Introduction}
Light beams possess both orbital angular momentum (OAM) and spin angular momentum (SAM). SAM is a consequence of the polarization of the light field, while OAM is associated with its spatial structure, specifically the azimuthal phase dependence. These spatially structured light beams are characterized by the topological charge (TC) or, equivalently, by the OAM they carry, which defines the number of $2\pi$ phase shifts along the azimuthal coordinate of the light beams. Allen et al.~\cite{Allen:92} were the first to demonstrate that Laguerre-Gauss (LG) beams, one of the solutions to the paraxial Helmholtz equation, carry OAM. Subsequently, various optical fields, including Bessel-Gauss (BG) and Mathieu beams, have been identified as carriers of OAM. Light beams with OAM are routinely generated in the visible and infrared spectral regimes using diverse tools and techniques such as spatial light modulators (SLMs)~\cite{Peng:20}, spiral phase plates (SPPs)~\cite{Chen:09}, mode converters~\cite{COURTIAL:99}, dielectric wedges~\cite{Izdebskaya:05}, deformable mirrors~\cite{Tyson:08}, metasurfaces~\cite{Ahmed:22}, among others. These OAM-carrying beams find applications in diverse areas, such as particle trapping and manipulation~\cite{padgett:11}, optical communication~\cite{wang:12}, microscopy and imaging~\cite{Otomo:14,Bala:22}, and detecting molecular chirality~\cite{forbes:19,ward:16}.

The superposition of optical beams carrying OAM exhibits intriguing properties with significant practical applications. Over the last three decades, numerous investigations have focused on the superposition of both OAM-carrying and non-OAM-carrying beams. Orlov et al.~\cite{ORLOV:03} explored the propagation of a composite beam consisting of two coaxial BG and LG vortex beams in free space. Their study highlighted the pivotal role of the propagation length and the ratio of amplitudes in influencing the vortical properties of the composite light field. In 1993, Basistiy et al.~\cite{BASISTIY:93} demonstrated that adding a small coherent background, in the form of a low-intensity plane wave, divides an initially multiple-charged vortex (of order $l$) into $l$ single-charged vortices. %They also showcased the formation of a second-order vortex beam through the nonlinear optical transformation of a first-order vortex beam using second harmonic generation (SHG)%. 
An investigation into the superposition of coaxial LG beams revealed a strong coupling between the number of vortices and their total TC, dependent on the relative widths and amplitudes of the input fields~\cite{Soskin:97}. The superposition of two coaxial BG beams of the same size in free space was also demonstrated~\cite{ORLOV:02}. In the near field, the composite beam exhibited a richer vortex content than the individual beams, evolving dynamically into a beam with a simple vortical structure while preserving the net TC during propagation.

The superposition of two non-coaxial LG beams was explored as well, revealing that the combined beam possesses a rich vortex content~\cite{Pyragaite:07}. Here, it was demonstrated that the locations of vortex cores in the combined beam depend not only on the separation between the beams but also on the introduced phase shift, TC, and their relative amplitudes. Pyragaite et al.~\cite{PYRAGAITE2003} delved into the interference of two LG beams for various intersecting angles, finding that the positions of vortex cores and the net TC content depend on the TCs of the individual input beams, propagation length, amplitude ratios, and the angle of intersection. A larger intersection angle resulted in a more intricate vortex content, showcasing different vortex interaction phenomena such as the pulling and pushing of vortices, dynamic inversion of the TC, and the nucleation of vortex pairs.

All the aforementioned studies predominantly focused on the superposition of LG, BG, or a combination of both. In the case of LG and BG beams, the size of the vortex core is tightly linked to the vortex charge, imposing a limitation on the utilization of these beam profiles for applications requiring a small core size with high vortex charge. This inherent constraint was overcome with the creation of a distinctive vortex beam profile known as the perfect optical vortex (POV) beam, characterized by a TC-independent core size and radial intensity distribution~\cite{ostrovsky:13}. POVs can be generated from a BG beam via an optical Fourier transformation~\cite{vaity:15} or using planar Pancharatnam-Berry phase elements~\cite{Liu:17}. Speckles created from a POV beam have a TC-independent size~\cite{Reddy:16} and interesting coherence properties~\cite{Vanitha:21}. POVs have found widespread use in various applications, including particle trapping in an optical tweezer~\cite{Chen:13}, cryptography~\cite{Pinnell:20,YangQing:22}, free-space optical communication~\cite{Shao:18}, and plasmonic structured illumination microscopy (PSIM)~\cite{Zhang:16}. In~\cite{Chen:13}, it was demonstrated that the utilization of the POV beam facilitated the rotation of a single microscopic particle, with a refractive index greater than that of its sorrounding medium, at a constant velocity around the annular intensity ring of the beam. This, in turn, overcomes the complexity of trapping a particle at intensity hot spots. They also pointed towards creating ring traps for atoms using POV beams. It is well-known that all OAM modes are orthogonal to each other. Therefore, the infinite OAM modes can be utilized to enhance the channel capacity, and spectral efficiency in a communication network i.e., all OAM modes can enable a single photon to carry unlimited amount of information. However, the utilization of LG modes in encoding unlimited amount of information in a single photon is problematic for various reasons: (1) the radial intensity distribution as well as the radius of the LG modes are strongly coupled to the TC of the beam, as a result of which, multiple LG modes can not be coupled into a communication system such as a fiber-optic network with a fixed annular index profile, (2) there is a trade-off between generating higher order LG modes from the spatial light modulator with their energy efficiency and quality, and (3) the side lobes of the LG modes also create undesirable effects in practical applications which has to be compensated as well. In~\cite{Shao:18}, authors tackled these issues by implementing the POV beam for free-space optical communication (FSO). Their results show that the use of POV beams multiplexing in a FSO link enhances the performance of the system, and lowers the bit error rate. In~\cite{Zhang:16}, authors reported that the inclusion of POV beams, instead of ordinary vortex beams such as the LG beams, to a PSIM system provided certain advantages, namely (1) an enhancement of the surface plasmon excitation efficiency nearly by 6.2 times as compared to the LG beam, and (2) the imaging resolution of the system went well below $200$~nm. A notable feature of the POV beam is its adaptability to scenarios where precise control over the core size and vortex charge is crucial. This prompts an intriguing avenue of exploration: investigating the interference of high-order POV beams separated by varying axial distances. Such a study holds promise for expanding our understanding of the interference characteristics of these unique beams and may uncover novel applications in diverse fields.

In this contribution, we investigate the superposition of two high-order POV beams, positioned at some axial distance apart. Our study demonstrates that both the axial distance and the phase shift between the two beams exert a strong influence on the resulting interference pattern. We illustrate a transformative process, wherein the interference pattern evolves from a single-ring intensity mode (observed when the beams are perfectly coaxial) to a two-ring pattern separated by a specific distance (notably, when the axial separation is $\approx 6.4$ times the individual beam waist). This transition in the interference pattern highlights the intricate interplay between the axial distance and phase shift in shaping the resultant optical field. Furthermore, we draw a comparative analysis between the interference patterns of two higher-order POV beams and LG beams with equivalent parameters. Our observations reveal that the splitting of the composite beam into its individual components occurs more rapidly in the case of LG beams. This comparative analysis provides valuable insights into the distinct interference characteristics of high-order POV beams and their divergence from other beam profiles, particularly LG beams, underlining the potential applications and advantages of POV beams in various optical scenarios.

\section{Results and Discussion}

\begin{figure}[ht]
\centering
\includegraphics[width=0.6\linewidth]{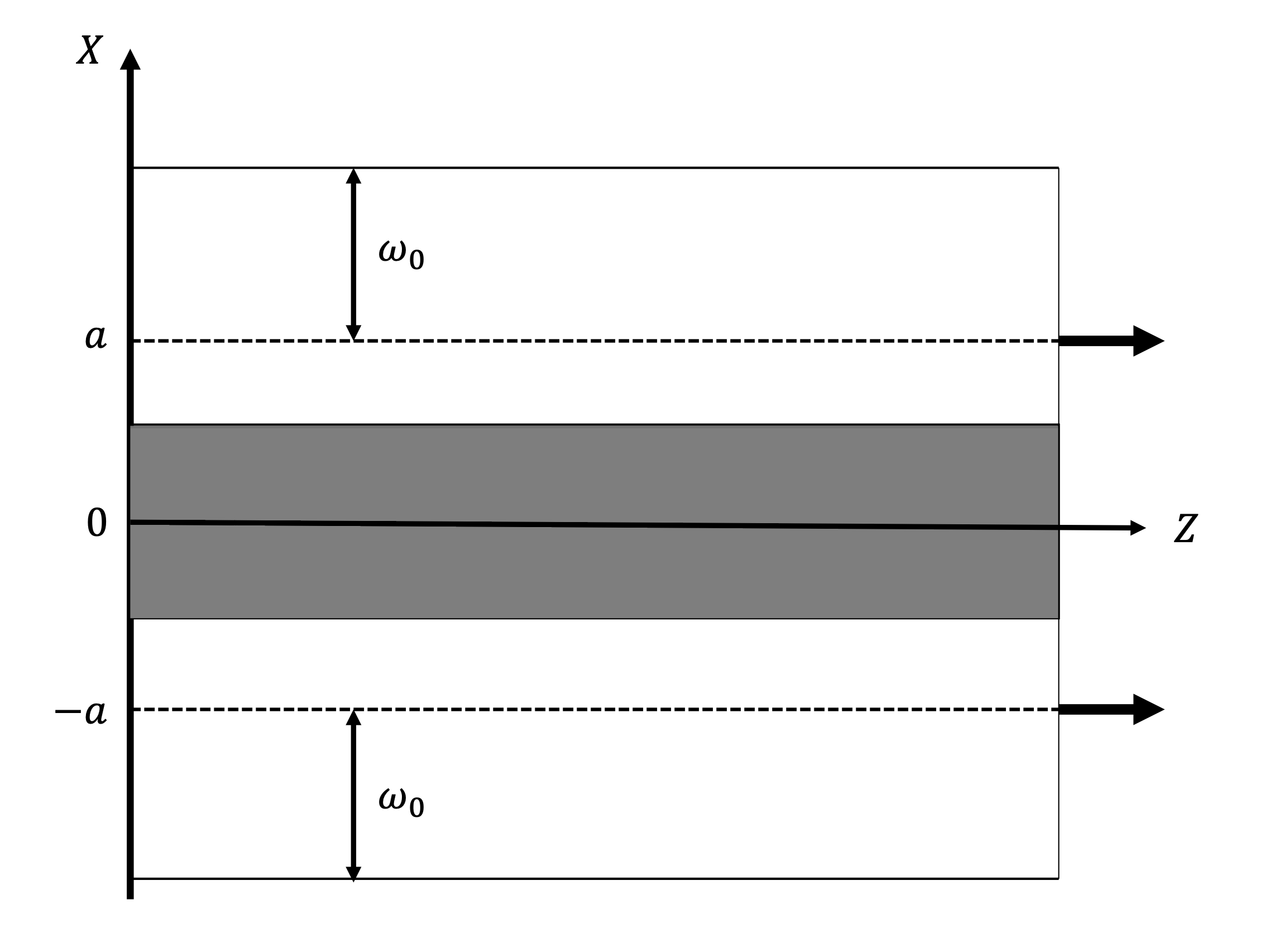}
\caption{Sketch of the overlapping of the two beams carrying OAM. Here, $2a$, and $\omega_{0}$ represent the inter-axial separation between the two beams, and individual beam waist, respectively.}
\label{fig1}
\end{figure}

A POV beam is characterized by a single, bright ring (in the transverse intensity profile) showcasing a TC-independent core size and radial intensity distribution. This beam is generated in the far field of a lens (i.e., at the focus), when a BG beam is supplied as an input. Therefore, the POV, and BG beams are considered as Fourier pairs. The complex field amplitude of the POV beam at the source plane, i.e., at $z=0$, is given as 
\begin{equation}
E(r,\theta)=E_0 \mathrm{e}^{-\frac{r^2}{\omega_0^2}}\mathrm{e}^{il\theta}I_l\left(\frac{2rR}{\omega_0^2} \right),    
\end{equation} 
where $r$ and $\theta$ are the polar coordinates. $E_0,R,\omega_0,$ and $l$ represent the constant field amplitude, the beam radius, the half-ring width (or Gaussian beam waist size), and the TC, respectively. Here, $I_l$ represents the modified Bessel function of the first kind of order $l$. In Cartesian coordinates, the complex field amplitude of the POV beam can be written as 
\begin{equation}
    E_{l0}(x,y)=E_0\frac{(x+iy)^l}{(x^2+y^2)^{l/2}}\mathrm{e}^{-\frac{x^2+y^2}{\omega_0^2}}I_l\left(\frac{2 R\sqrt{x^2+y^2}}{\omega_0^2} \right)\label{Field},
\end{equation} 
where $x=r \cos(\theta)$ and $y=r \sin(\theta)$, respectively. The vortex core is located at $x=0\, ,y=0$. In our work, we study the superposition of two high-order collinear identical POV beams, separated by a distance $2a$, namely the inter-axial distance between the two beams (see Fig.~\ref{fig1}). We termed $a$ as the axial distance between the beam and the optical axis . Importantly, we are not considering the propagation characteristics of the beams, and our theoretical analysis is fully devoted to the beams located at $z=0$ . This is due to the  fact that, unlike the LG beam, the POV beam is not propagation invariant, i.e. , the beam diffracts both towards, and away from its center, and eventually degrades into a Bessel-like beam after a small propagation distance~\cite{Bikash:24}. However, the invariance propagation nature of the beam can be exploited in fiber-optic communications.

In this scenario, the complex field amplitude of the composite beam can be written as 
\begin{equation}
E_l(x,y)=E_{l0}(x-a,y)+\mathrm{e}^{i \Phi}E_{l0}(x+a,y),
\label{interf}
\end{equation}
where $\Phi$ is the relative phase shift introduced between the two beams. To locate the vortex cores of the composite beam, we use two different approaches: (1) by solving $\mathrm{Re}[E_l(x,y)]=0$ and $\mathrm{Im}[E_l(x,y)]=0$ analytically and (2) by plotting the line intensity profile of the composite beam in both $x$, and $y$ dimensions. The results obtained from these approaches are consistent. In this section, we discuss the following cases : the interference of two high-order POV beams with (1) $l=2$, $\Phi=0$, $R=0.04$~m, (2)  $l=2$, $\Phi=\pi$, $R=0.04$~m, (3)  $l=3$, $\Phi=0$, $R=0.04$~m, (4) $l=2$, $\Phi=0$, $R=0.02$~m and the interference of two high-order LG beams with (5) $l=2$, $\Phi=0$, (6)  $l=2$, $\Phi=\pi$, (7)  $l=3$, $\Phi=0$.

\subsection{Case 1} \label{C1}

\begin{figure*}[ht]
\centering
\includegraphics[width=1\linewidth]{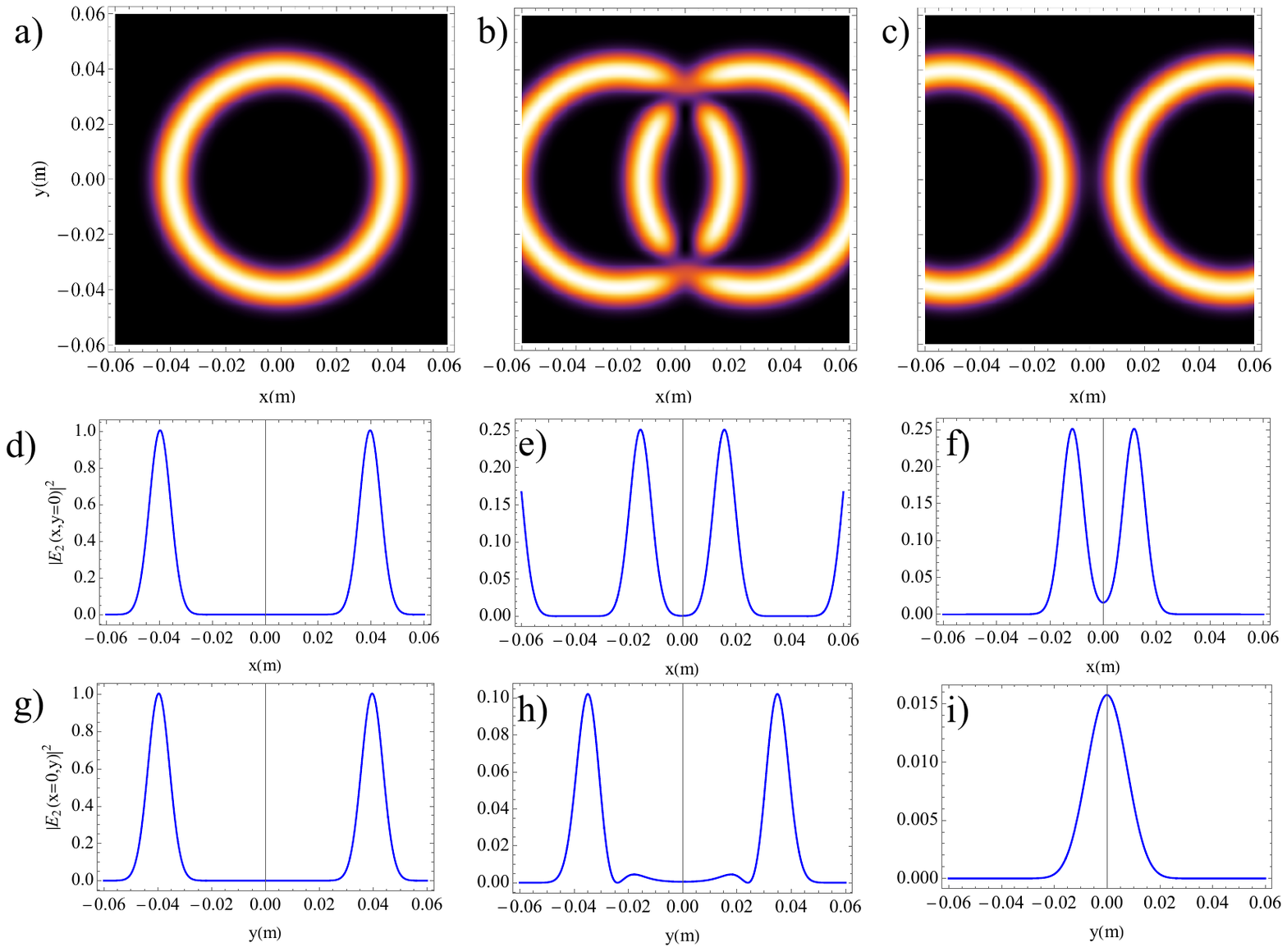}
\caption{Interference pattern of two identical high-order POV beams of $l=2$ and $\Phi=0$, for different values of the dimensionless beam separation $\beta$ ($\beta=\frac{a}{\omega_0}$). In (a) $\beta=$ 0, (b) $\beta=$ 3.0, (c) $\beta=$ 6.4 (see Visualization 1). Their corresponding line intensity plots along $x$, and $y$ dimensions are shown in (d)-(f), and (g)-(i), respectively. The simulation parameters used here are: $\omega_0=0.008$~m, $R=0.04$~m.}
\label{F2}
\end{figure*}

In this case, the complex field amplitude of the composite POV beam can be written by substituting Eq.~\ref{Field} into Eq.~\ref{interf} for $l=2$, $\Phi=0$, $R=0.04$~m, and $\omega_{0}=0.008$~m. The transverse intensity distribution ($|E_2 (x,y)|^2$ ) of the composite beam for different values of the axial separation distance is depicted in Figs.~\ref{F2} (a) to (c). It can be clearly seen that when the axial separation between the two POV beams is set to zero, i.e., a perfect overlapping of the two input POV beams (Fig.~\ref{F2}(a)), a single bright ring in the transverse intensity profile of the composite POV beam is formed. The amplitude of the generated composite beam is twice the amplitude of the either input POV beams. This explains a perfect constructive interference between the individual beams. Here, the size and width of the composite beam remain unchanged. As we keep on increasing the inter-axial separation between the beams, deviations from the ideal intensity distribution are observed (Figs.~\ref{F2}(b) and (c), and for complete details see Visualization 1). There also occurs a shifting in the vortex cores accompanied by a change in the beam intensity. This implies that the size of the dark core reduces in the $x$ dimension, as we increase the inter-axial separation. However, the topological charge of the composite beam remains the same i.e., +2. It can also be observed from the line intensity profiles shown in Figs.~\ref{F2}(d)-(i). A net $l=+2$ is well-maintained for the composite beam across all stages of the axial separation. When we set $\beta(=\frac{a}{\omega_{0}})\approx6.4$, there is an apparent splitting of the composite POV beam into two distinct POV beams of $l=+2$ each. The location of the vortex cores in the composite beam can be found either by (1) considering $\mathrm{Re}[E_2(x,y)]=0$ and $\mathrm{Im}[E_2(x,y)]=0$ or (2) by plotting the line intensity profile of the composite POV beam for different values of $\beta$. We plotted the normalized line intensity profile of the composite POV beam both in $x$, and $y$ dimensions given by Eq.~\ref{interf}. The position of the vortex cores can be obtained by considering the locations along $x$, and $y$, where the intensity vanishes or is vanishingly small (that essentially means the field amplitude also vanishes at these locations). The line plots of the composite beam for different values of the dimensionless parameter $\beta$ are shown in Figs.~\ref{F2} (d) to (i). At $\beta=0$, a composite POV beam is formed, due to the perfect overlapping (i.e., there exists a constructive interference) of the two identical POV beams, with an amplitude twice that of the amplitude of the either POV beam. Therefore, the intensity of the composite beam is maximum at this inter-axial separation, as shown in Fig.~\ref{F2}(d,g). As the inter-axial separation between the two beams increases, there occurs many interesting effects: (1) there is a decrease in the composite beam intensity, and (2) the vortex dynamics also changes as can be seen from Figs.~\ref{F2}(e,h) and Figs.~\ref{F2} (f,i). We also observe a splitting of the composite POV beam into two distinct POV beams of TC $l=+2$ each at $\beta=6.4$ (see Visualization 1). However, the net TC content of the composite beam remains +2 at the different stages. For instance, at $\beta=6.4$, two POV beams of TC, $l=+2$, are formed accompanied by two vortices with TC, $l=-1$, which are always present in the region where the field amplitude is vanishingly small. This, in turn , makes the total TC, $l=+2$ at this stage. In this article, we have prioritized the numerical approach to locate the vortex cores in the case of composite POV beams, whereas, both analytical, and numerical approaches are used in the case of composite LG beams (see next sections).

\subsection{Case 2}
\begin{figure*}[ht]
\centering
\includegraphics[width=1\linewidth]{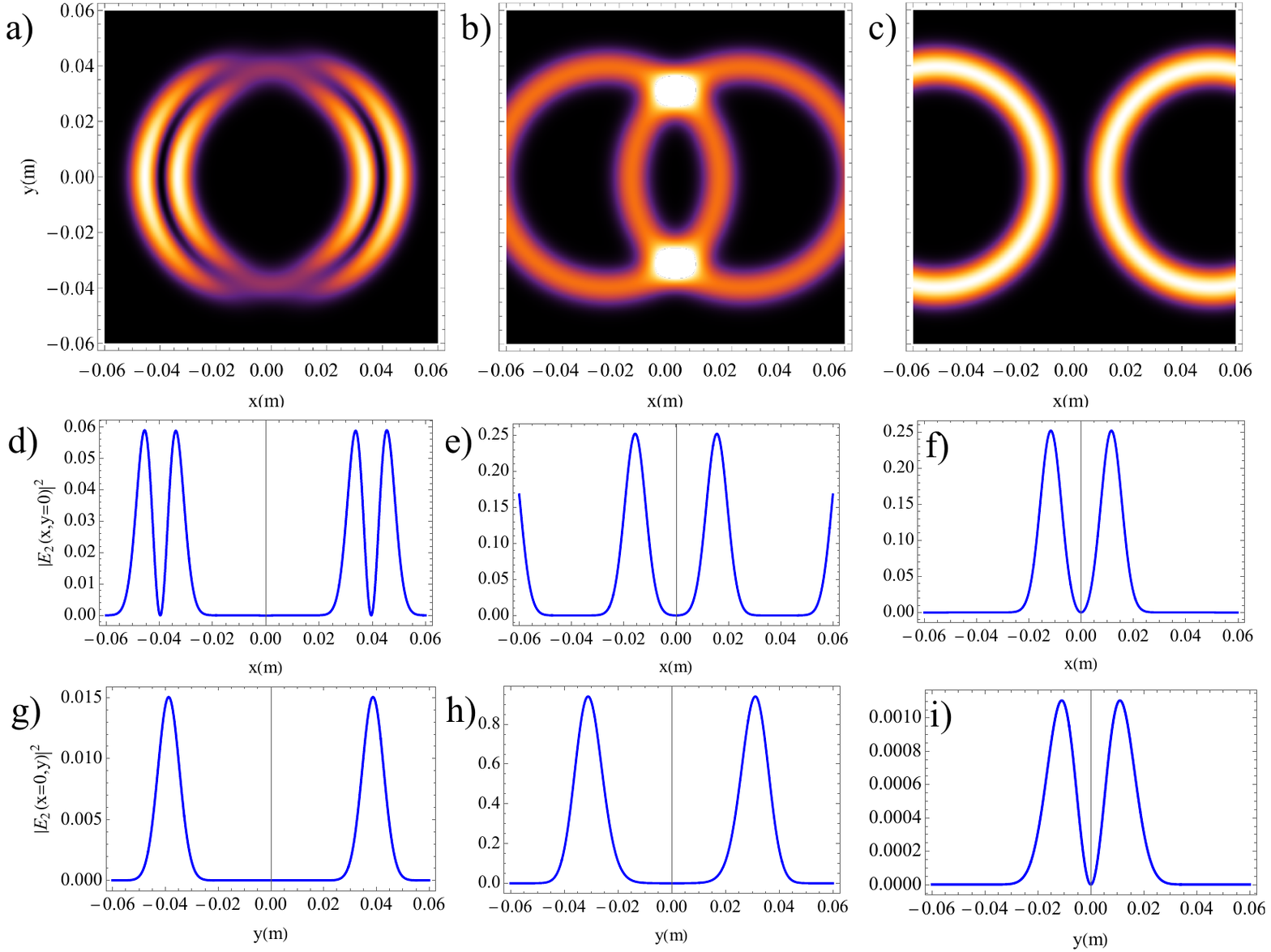}
\caption{Interference pattern of two identical high-order POV beams of $l=2$ and $\Phi=\pi$, for different values of the dimensionless beam separation $\beta$ . In (a) $\beta=$ 0.3, (b) $\beta=$ 3.0, (c) $\beta=$ 6.4 (see Visualization 2). Their corresponding line intensity plots along $x$, and $y$ dimensions are shown in (d)-(f), and (g)-(i), respectively. The simulation parameters used here are: $\omega_0=0.008$~m, $R=0.04$~m.}
\label{F3}
\end{figure*}

For the second case, we set the values $l=2$, $\Phi=\pi$, $R=0.04$~m. This choice of parameters is justified by the fact that the vortical structure of the composite beam is sensitive to changes in the relative phase shift of the beams, $\Phi$. Once again, the complex field amplitude of the composite POV beam can be extracted from Eq.~\ref{interf}. The simulated transverse intensity distribution of the composite POV beam is shown in Fig.~\ref{F3}. From the figure, it is clear that by introducing this relative phase shift, one can completely alter the transverse intensity distribution of the composite beam i.e., the interference pattern. It can be seen from Fig.~\ref{F3}(a) that, the degree of destructive interference is maximized at $\beta=0.3$. However, it is crucial to note that, despite these modifications, a consistent net $l=+2$ is maintained for larger values of $\beta$ (see Visualization 2). The dynamics of the vortices in this scenario differ from the results presented in  sub-section \ref{C1}. There also occurs a change in the composite beam intensity as the value of $\beta$ changes. Nonetheless, in a similar fashion, we found that the composite beam splits into two POV beams, each with $l=+2$, when $\beta=6.4$ i.e., the axial separation is approximately 6.4 times the input beam waist of the POV beams (see Visualization 2). In Figs.~\ref{F3}(d)-(i), we show the location of the vortex cores for the interference patterns shown in Figs.~\ref{F3} (a) to (c). The core values are found by plotting the line intensity profile of the composite beam along both $x$ and $y$-dimensions as shown in Figs.~\ref{F3}(d)-(e), and (g)-(i), respectively. It can be clearly seen from Figs.~\ref{F3}(d,g) that the beam intensity is minimum at $\beta=0.3$, which mimics the destructive interference due a relative phase shift of $\pi$ between the two beams. As the inter-axial separation increases further, the composite beam intensity also increases (see Visualization 2). However, a topological charge, $l=+2$, is maintained across all stages. At $\beta=6.4$, the composite POV beam splits into two distinct POV beams of $l=+2$ each (see Figs.~\ref{F3}(f,i)). 

\subsection{Case 3}

\begin{figure}[htbp]
\centering
\includegraphics[width=1\linewidth]{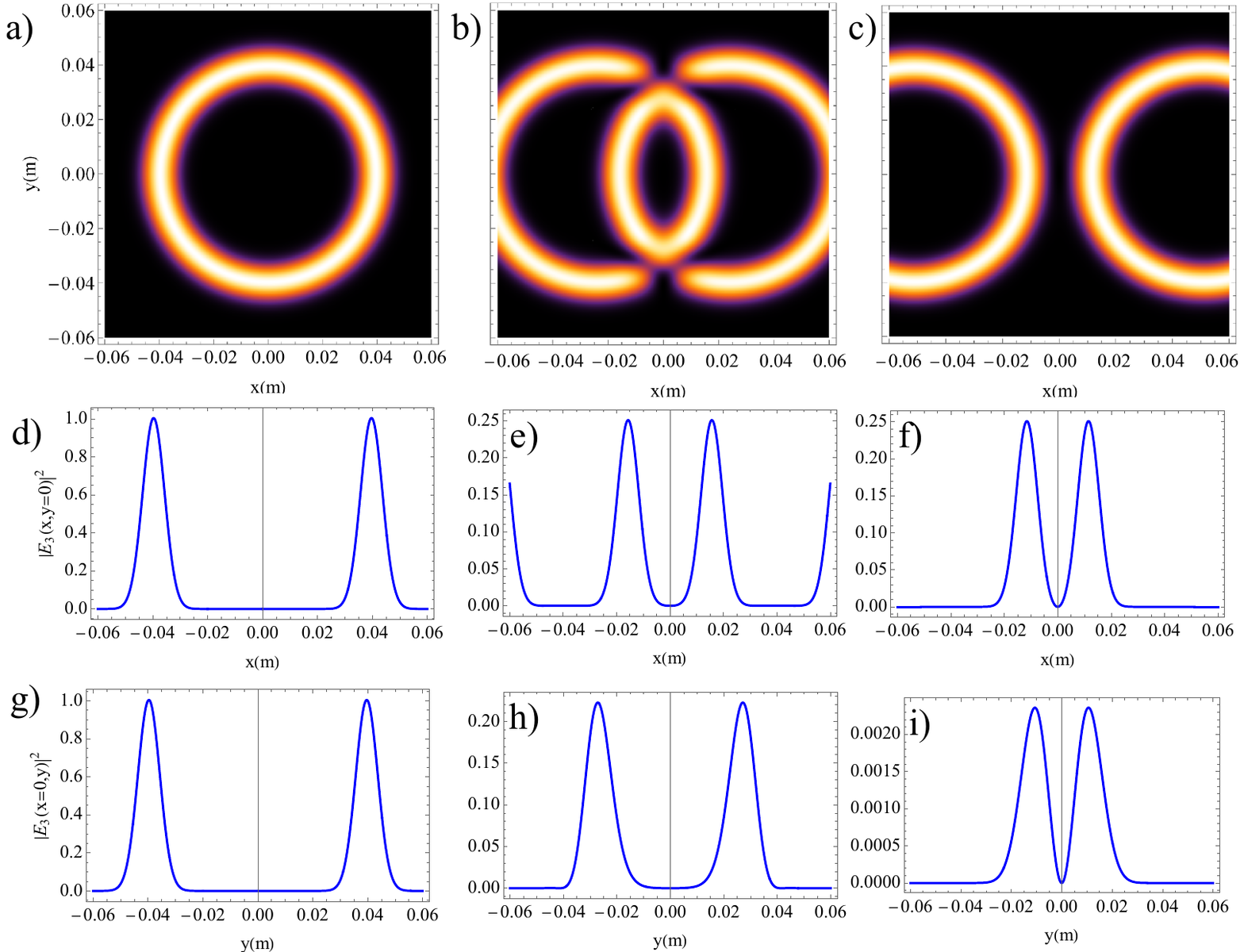}
\caption{Interference pattern of two identical high-order POV beams of $l=3$ and $\Phi=0$, for different values of $\beta$ ($\beta=\frac{a}{\omega_0}$). In (a) $\beta=$ 0, (b) $\beta=$ 3.0, (c) $\beta=$ 6.4 (see Visualization 3). Their corresponding line intensity plots along $x$, and $y$ dimensions are shown in (d)-(f), and (g)-(i), respectively . The simulation parameters used here are: $\omega_0=0.008$~m, $R=0.04$~m.}
\label{F4}
\end{figure}

Next, we study the case for $l=3$, $\Phi=0$ and $R=0.04$~m. This case is important to elucidate how increasing the topological charge, $l$, of the input beams affects the transverse intensity distribution of the composite beam. The resulting transverse intensity distribution ($|E_3 (x,y)|^2$ ) of the composite beam for different values of the inter-axial separation is depicted in Fig.~\ref{F4}. We maintain identical beam parameters as in \textit{Case 1}, and the simulations are conducted over the same inter-axial separations. We observe that the interference pattern begins to diverge from the initial case at $\beta=2$ (see Visualization 3). However, the overall TC content of the composite beam remains at $l=+3$. Importantly, there is no discernible difference in the transverse intensity profiles between the composite beams with $l=+3$ and $l=+2$ at $\beta=0$ (see Fig.~\ref{F2}(a), and Fig.~\ref{F4}(a)). This characteristic distinguishes a POV beam from all other optical beams carrying OAM i.e., the size of the dark core, and the radial intensity distribution remain unchanged with an increase in the TC. At an axial separation of $\beta=6.4$, the composite beam bifurcates into two POV beams with $l=+3$ as shown in Fig.~\ref{F4}(c). 

The calculated line intensity profiles along $x$ and $y$-dimensions are shown in Figs.~\ref{F4} (d)-(f), and (g)-(i), respectively. From these figures, one can extract the changes in the vortex core position. At $\beta=0.3$, a single vortex of TC, $l=+3$, is located at $x=0, y=0$ as depicted in Figs.~\ref{F4}(d,g). As we change the value of $\beta$, the core dynamics changes significantly as evident from Figs.~\ref{F4}(e,h)-(f,i). When we set $\beta=6.4$, two POV beams with TC, $l=+3$ each, are formed along with three negatively charged vortices of TC $-1$ each in the region where the field amplitude, hence the intensity, is zero (see Figs.~\ref{F4}(f,i). The  case of $\beta=6.4$, as depicted in Fig.~\ref{F4}(f,i), holds particular significance for analyzing the interference of all high-order POV beams and can be deemed a "critical" value. It is noteworthy that this value remains constant for the beam parameters utilized in our simulations.

\subsection{Case 4}

\begin{figure}[htbp]
\centering
\includegraphics[width=1\textwidth]{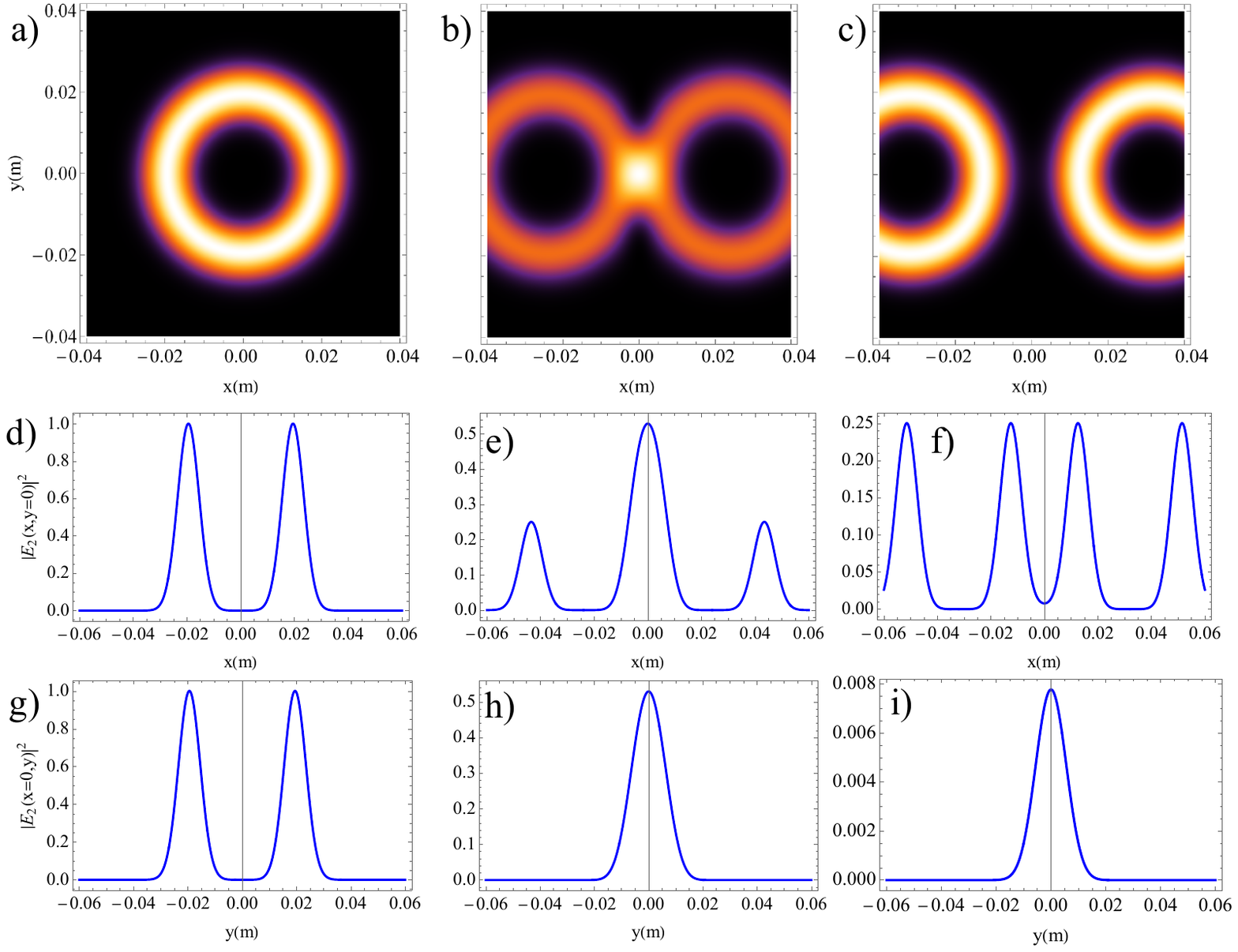}
\caption{Normalized line intensity profile of the composite POV beam of TC, $l=2$, and $\Phi=0$. In (a) $\beta=$ 0, (b) $\beta=$ 3.0, (c) $\beta=$ 4.0 (see Visualization 4). Their corresponding line intensity plots along $x$, and $y$ dimensions are shown in (d)-(f), and (g)-(i), respectively. The simulation parameters used here are: $\omega_0=0.008$~m, $R=0.02$~m.}
\label{F5}
\end{figure}

Considering the complex field amplitude of the composite POV beam, in addition to $l$ and $\Phi$, we have another degree of freedom that can modify the interference of the two POV beams: the beam radius. In this context, we investigate the impact of altering (specifically reducing) the POV beam radius on the interference pattern. For this case, we maintain $l=2$, $\Phi=0$, and $\omega_0=0.008$ m as in \textit{Case 1}, but we vary the beam radius to $R=0.02$ m. The transverse intensity distribution ($|E_2 (x,y)|^2$ ) of the composite beam for different values of the inter-axial separation distance is depicted in Fig.~\ref{F5}.  When the two beams perfectly overlap, they form a single-ring composite POV beam with TC, $l=+2$, as shown in Fig.~\ref{F5}(a). However, deviations from the ideal profile emerge as the axial separation between the beams changes. It is observed that the nature of the interference pattern in \textit{Case 1} and \textit{Case 4} remains the same, i.e. there is a change in the composite beam width along both $x$ and $y$ dimensions accompanied by a change in the beam intensity, for $\beta=0.6$ and $0.74$ (see Fig.~\ref{F2} an Visualization 1 and Fig.~\ref{F5} and Visualization 4). However, the TC of the composite beam remains $+2$ across these stages. There is a splitting of the vortex core of TC $l=+2$ into two cores of TC $l=+1$, each at $\beta=1.2$. A net $l=+2$ is maintained in the composite beam across all axial separations, although the vortex dynamics is altered. Particularly, at $\beta=4$, the composite beam distinctly splits into two identical POV beams, each with $l=+2$. It is worth noting that by reducing the beam radius, the axial separation at which the splitting occurs decreases significantly. It is also important to highlight that, as the beam POV beam radius decreases further, the splitting of the composite POV beam occurs more rapidly i.e., at an extremely small inter-axial separation like the composite LG beams. This effect can be seen from the line intensity plots depicted in the second and third row of Fig.~\ref{F5}, for different values of $\beta$. At $\beta=0$, a vortex with TC, $l=+2$, is located at $x=0, y=0$ (see Figs.~\ref{F5}(d,g)). As we keep on increasing the value of the inter-axial separation, the dynamics of the vortices changes. At $\beta=3.0$, two vortices with TC, $l=+2$ each, are formed (see  Fig.~\ref{F5}(e)). There also exists two vortices with charges, $l=-1$ each, in the region where the field amplitude is vanishingly small as shown in  Fig.~\ref{F5}(h). This, in turn, makes the total TC, $l=+2$, at this stage. Again at $\beta=4.0$, the composite POV beam decomposes into two distinct POV beams of TC, $l=+2$ each, with two negatively charged vortices located in the region of zero field amplitude (see  Fig.~\ref{F5}(f,g)).

To gain deeper insights into the vortex dynamics within composite beams carrying OAM, we contrast the interference patterns of high-order POV beams with those of high-order LG beams, using identical sets of beam parameters, including beam waist, amplitudes, and TC.

\subsection{Case 5}
\begin{figure}[h!]
\centering
\includegraphics[width=1\linewidth]{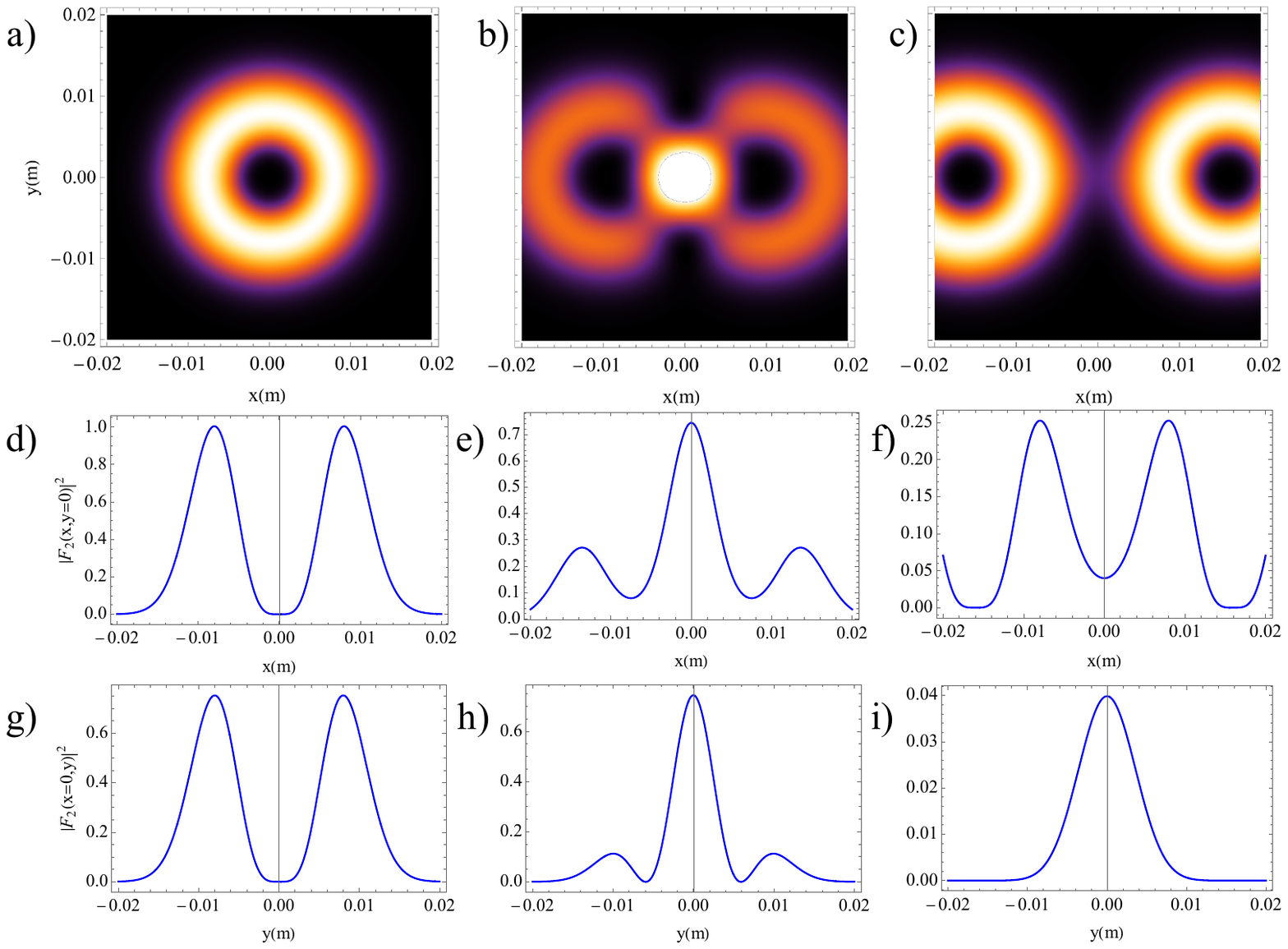}
\caption{Interference pattern of two identical high-order LG beams of $l=2$ and $\Phi=0$, for different values of $\beta$. In (a) $\beta=$ 0, (b) $\beta=$ 0.74, (c) $\beta=$ 2.0 (see Visualization 5). Their corresponding line intensity plots along $x$, and $y$ dimensions are shown in (d)-(f), and (g)-(i), respectively. Here we set $\omega_0=0.008$~m.}
\label{F6}
\end{figure}

The complex field amplitude of the composite LG beam of TC $l=2$, and $\Phi=0$ can be written as:

\begin{eqnarray}
F_2(x,y)=F_0\Big(\frac{(x-a)+iy}{\omega_0}\Big)^2e^{-\frac{(x-a)^2+y^2}{\omega_0^2}}+F_0\Big(\frac{(x+a)+iy}{\omega_0}\Big)^2e^{-\frac{(x+a)^2+y^2}{\omega_0^2}}.
\end{eqnarray}
The calculated intensity distribution ($|F_2 (x,y)|^2$ ) of the composite LG beam of TC +2, and $\Phi=0$ for different values of $\beta$ are presented in Fig.~\ref{F6} , where $\beta=\frac{a}{\omega_0}$, $a$ is the axial separation distance of the beam from the optical axis, and $\omega_0$ is the beam waist of the Gaussian beam. We observe the merging of two LG beams into a single LG composite beam with $l=+2$ when $\beta=0$ as shown in Fig.~\ref{F6}(a). However, as the axial separation increases, deviations in the interference pattern emerge as depicted in Figs.~\ref{F6}(b)-(c) (see Visualization 5 for more details). Nonetheless, the net TC consistently remains $l=+2$. With a continued increase in the axial separation, two well-separated LG beams, each with $l=+2$, are formed as shown in Fig.~\ref{F6}(c). The locations of the vortex cores of the composite LG beam can be found either (1) by analytically solving the equations $\mathrm{Re}[F_2(x,y)]=0$ and $\mathrm{Im}[F_2(x,y)]=0$ i.e.,
\begin{eqnarray}
    ((x-a)^2-y^2)e^{\frac{2xa}{\omega_0^2}}+((x+a)^2-y^2)e^{-\frac{2xa}{\omega_0^2}}=0 \label{xC5}
\end{eqnarray}
and
\begin{eqnarray}
    y(x-a)e^{\frac{2xa}{\omega_0^2}}+y(x+a)e^{-\frac{2xa}{\omega_0^2}}=0\label{yC5}
\end{eqnarray}
or (2) by numerically calculating the normalized line intensity profile of the composite LG beam along both $x$, and $y$ dimensions.

Let us consider the first case i.e., locating the vortex cores in the composite beam by solving Eqs.~\ref{xC5}, and \ref{yC5} analytically. For $y\neq0$, Eq.~\ref{yC5} can be re-written as: 

\begin{equation}
x(e^{\frac{2xa}{\omega_{0}^{2}}}+e^{-\frac{2xa}{\omega_{0}^{2}}})=a(e^{\frac{2xa}{\omega_{0}^{2}}}-e^{-\frac{2xa}{\omega_{0}^{2}}}).
\end{equation}

The final expression for locating the position of the core at the x-axis and y-axis becomes: 
     
 \begin{equation}
     x=\tanh{\frac{2xa}{\omega_{0}^{2}}}
     \label{xsol}
 \end{equation}

and, 

 \begin{equation}
     y=\pm\sqrt{a^{2}-x^{2}}.
     \label{ysol}
 \end{equation}

At $\beta=0$ (i.e., $a=0$), two LG beams of TC $+2$ each overlap perfectly i.e., they interfere constructively, and a composite LG beam is formed with an amplitude twice that of the individual beam's amplitude. The vortex core of the resulting beam is located at $x=0, y=0$ as shown in Fig.~\ref{F6}(a) .  For $0<a<\frac{\omega_{0}}{\sqrt{2}}$ (i.e., $0<\beta<\frac{1}{\sqrt{2}}$), we get $x=0$, and $y=\pm a$. So, in this case, there exists two single charged vortices in the composite LG beam, and their cores are $2a$ distance apart, i.e., a vortex of TC $+1$ is located at $y=+a$, and the other one with TC $+1$ at $y=-a$,  as depicted by two dark cores in the y dimension (see Visualization 5 for details). The vortices change their sign i.e., from $+1$ to $-1$, and each of them are nucleated at the origin of each vortex to give rise to six vortices in total, when $a=\frac{\omega_{0}}{\sqrt{2}}$ (i.e., $\beta=\frac{1}{\sqrt{2}}$). Their distribution can easily be seen in the Visualization 5. However, a net TC of +2 is maintained at this stage as well. For $a>\frac{\omega_{0}}{\sqrt{2}}$ (i.e., $\beta>\frac{1}{\sqrt{2}}$), there exists six vortices in the composite LG beam, and their cores are located at the circle given by the equation $x^{2}+y^{2}=a^{2}$ (see Visualization 5 for more details). It is important to highlight that a net TC of +2 is maintained at different stages. When $a>>\omega_{0}$ (i.e., $\beta>>1$), the solutions of Eqs.~\ref{xsol} and \ref{ysol} are $x=0, y=\pm a$, and $x=\pm a, y=0$. As a result, two distinct LG beams of TC +2 each are formed with the vortex cores located at $(a,0)$, and $(-a,0)$ as shown in Fig.~\ref{F6}(c). It is also important to mention that two vortices of TC $-1$ are located in the region in between the two LG beams, where the field amplitude is very small as can be seen in Fig.~\ref{F6}(c).

Now, let's consider the second method, i.e., to analyze a line intensity plot of the composite beam. Thus, we can explore uts vortex dynamics. The normalized line intensity profiles of the composite beam of TC, $l=2$, and $\phi=0$ are shown in Figs~\ref{F6}(d)-(i). It can be seen that at $\beta=0$, as shown in Fig.~\ref{F6}(d), the intensity of the composite LG beam is maximum, as it is formed due to the constructive interference of the two LG beams. The vortex core is located at $x=0$, and  $y=0$ as depicted in Figs.~\ref{F6}(d,g). As the inter-axial separation between the two beams increases, there is a change (i.e., decrease) in the beam intensity as well as the location of the vortex cores as evident from Figs.~\ref{F6}(e)-(i). At $\beta=2$, the composite LG beam splits into two distinct LG beams of TC +2 each. Furthermore, two vortices of charge $-1$ are always present in between the two well separated beam, where the field amplitude vanishes. This is what makes the TC of the composite LG beam $+2$ at this stage as shown in Figs.~\ref{F6}(f,i).

\subsection{Case 6}

\begin{figure}[htbp]
\centering
\includegraphics[width=1\linewidth]{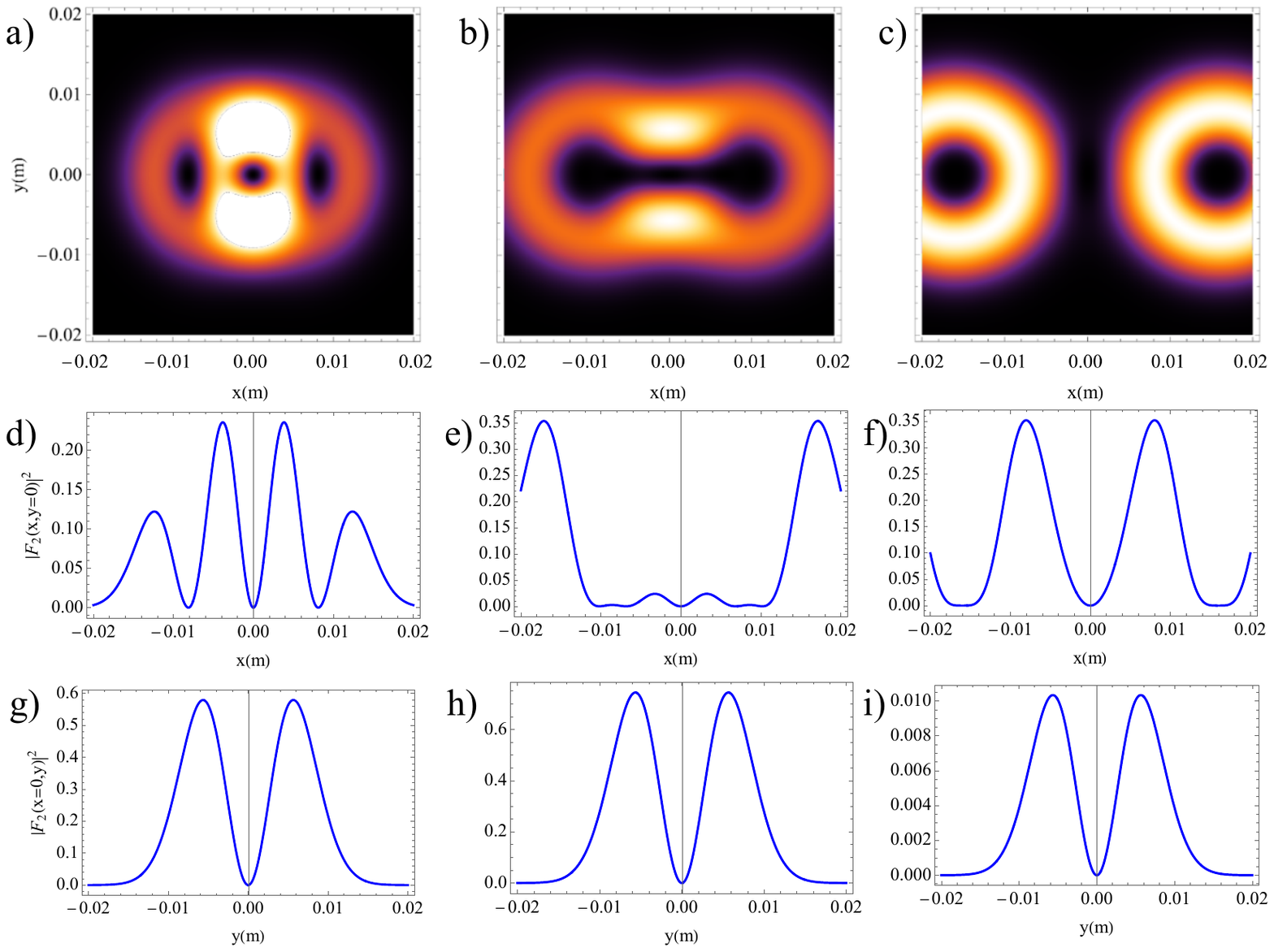}
\caption{Interference pattern of two identical high-order LG beams of $l=2$ and $\Phi=\pi$, for different values of $\beta$ . In (a) $\beta=$ 0.3, (b) $\beta=$ 1.14, (c) $\beta=$ 2.0 (see Visualization 6). Their corresponding line intensity plots along $x$, and $y$ dimensions are shown in (d)-(f), and (g)-(i), respectively. In all the cases $\omega_0=0.008$~m.}
\label{F7}
\end{figure}

The complex field amplitude of the composite LG beam of TC $l=2$, and $\Phi=\pi$ can be written as:
\begin{eqnarray}
F_2(x,y)=F_0\Big(\frac{(x-a)+iy}{\omega_0}\Big)^2e^{-\frac{(x-a)^2+y^2}{\omega_0^2}}-F_0\Big(\frac{(x+a)+iy}{\omega_0}\Big)^2e^{-\frac{(x+a)^2+y^2}{\omega_0^2}}
\end{eqnarray}
The numerically calculated intensity distribution ($|F_2 (x,y)|^2$ ) of the composite LG beam of TC 2, and $\Phi=\pi$ for different values of $\beta$ are shown in Fig.~\ref{F7} (see Visualization 6). The locations of the vortex cores of the composite LG beam can be found analytically by solving the equations $\mathrm{Re}[F_2(x,y)]=0$ and $\mathrm{Im}[F_2(x,y)]=0$ analytically or by the numerical approach discussed earlier. Let's discuss the analytical approach for this case.

The real and imaginary parts of $F_2(x,y)$ gives:
\begin{eqnarray}
    ((x-a)^{2}-y^{2}) e^{\frac{2xa}{\omega_{0}^{2}}}-((x+a)^{2}-y^{2}) e^{-\frac{2xa}{\omega_{0}^{2}}}=0\label{xC11}
\end{eqnarray}
and
\begin{eqnarray}
     y(x-a)e^{\frac{2xa}{\omega_0^2}}-y(x+a)e^{-\frac{2xa}{\omega_0^2}}=0\label{yC12}
\end{eqnarray}
Equations~\ref{xC11}  and ~\ref{yC12} can only be solved for $y=0$, e.g. Eq.~\ref{xC11} results

\begin{eqnarray}
    (x^{2}+a^{2})\tanh{\frac{2xa}{\omega_0^2}}=2ax\label{xC13}.
\end{eqnarray}
For $a\neq 0$, Eq.~\ref{xC13} is valid only when $x=0$. The vortex cores are also located at $x\neq 0$. When $\beta(=\frac{a}{\omega_{0}})<<1$, $x=\omega_{0}$. Therefore, at $y=0$, three vortices of TC +1 are present (three dark regions for three different values of $x$, and $y=0$) as shown in Fig.~\ref{F7}(a). Likewise, the net TC in this stage is +3. At $\beta=1.02$, the vortex located at $x=0$ changes its TC from +1 to -1 accompanied by the formation of two more vortices of TCs +1 (see Visualization 6 for details). Two more vortices of TCs +1 each are also located at $x\neq0$. If the value of $\beta$ is increased further, the collision of vortices occurs and this results in the formation of doubly charged vortices as shown in Figs.~\ref{F7}(b),(c). Note that, one vortex with TC $l=-1$ is also located in the region between the two LG beams (at $x=0$) of TCs +2 each as the intensity is very small there. That is the reason why a net TC of +3 is maintained in each stage. The numerical method, described earlier, can also be used to validate this analytical explanation. The numerically computed line intensity plots for three different values of $\beta$ are shown in Fig.~\ref{F7}(d)-(i). At $\beta=0.3$, three vortices of TC, $l=+1$ each, are located in the composite beam, which can be clearly seen from Figs.~\ref{F7}(d,g). Therefore, the net TC at this stage is $+3$. As we increase the value of $\beta$ further, vortex dynamics such as nucleation, and collision of vortices occurs. At $\beta=2.0$, two well-separated LG beams of TC, $l=+2$ each, are formed. It is also important to highlight that there exists a vortex of TC $l=-1$ at $x=0,y=0$, which makes the overall TC $+3$ at this stage (see Fig.~\ref{F7}(f,i)).

\subsection{Case 7}

Further investigation involves examining the effect of changing the $l$ of the LG beams on the interference pattern. For this analysis, we maintain the same beam waist and amplitudes but change $l$ to $l=3$ and $\Phi=0$.
Finally, the complex field amplitude of the composite LG beam of TC 3, and $\Phi=0$ can be written as
\begin{eqnarray}
F_3(x,y)=F_0\Big(\frac{(x-a)+iy}{\omega_0}\Big)^3e^{-\frac{(x-a)^2+y^2}{\omega_0^2}}+F_0\Big(\frac{(x+a)+iy}{\omega_0}\Big)^3e^{-\frac{(x+a)^2+y^2}{\omega_0^2}}\label{xC14}.
\end{eqnarray}

\begin{figure}[htbp]
\centering
\includegraphics[width=1\textwidth]{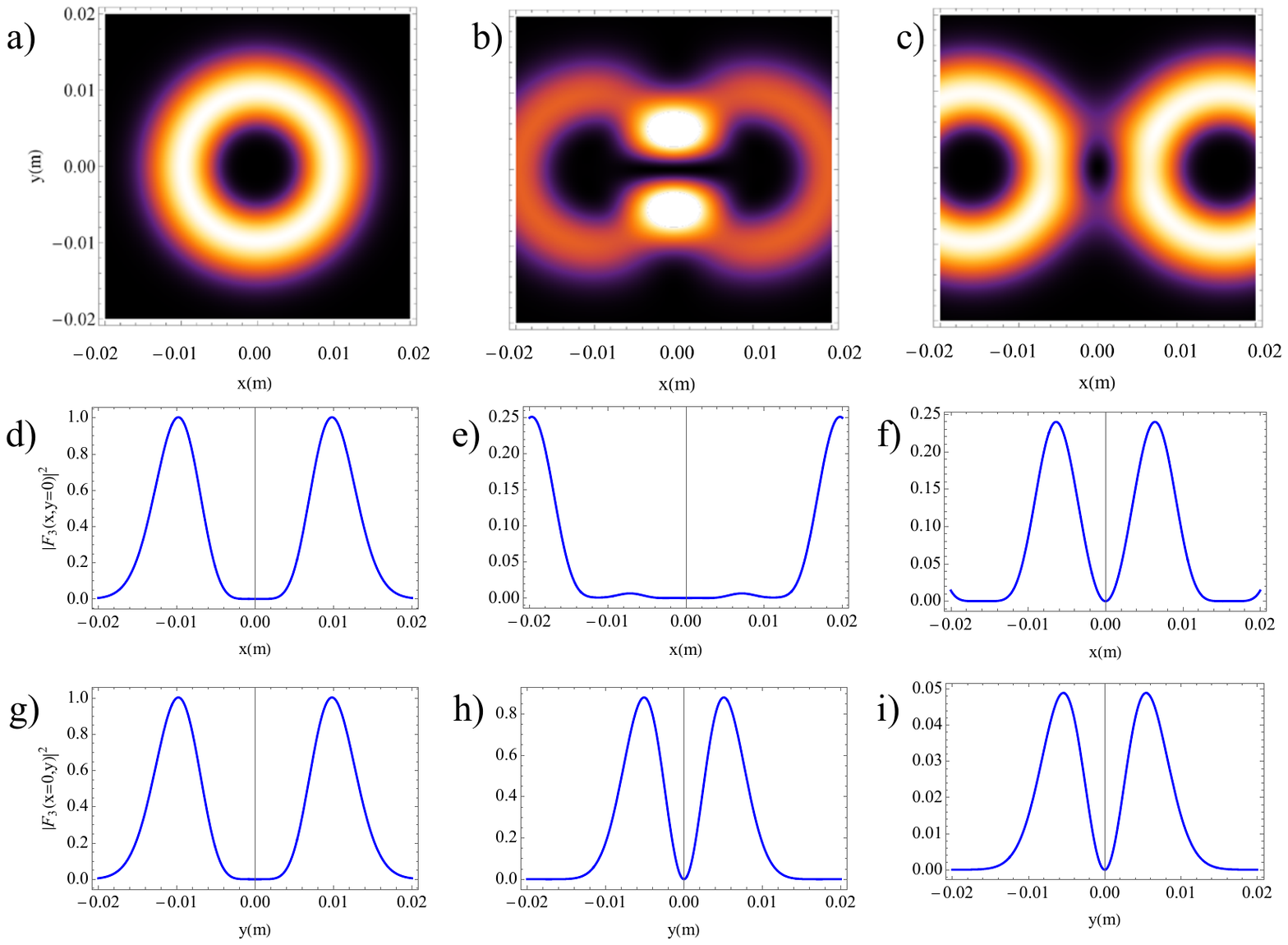}
\caption{Normalized line intensity profile of the composite LG beam of TC, $l=3$, and $\Phi=0$. In (a) $\beta=$ 0, (b) $\beta=$ 1.24, (c) $\beta=$ 2.0 (see Visualization 7). Their corresponding line intensity plots along $x$, and $y$ dimensions are shown in (d)-(f), and (g)-(i), respectively. The simulation parameters used here are: $\omega_0=0.008$~m.}
\label{F8}
\end{figure}

The numerically calculated intensity distribution ($|F_3 (x,y)|^2$ ) of the composite LG beam of TC 3, and $\Phi=0$ for different values of $\beta$ are shown in Fig.~\ref{F8}.

The results reveal that when the two LG beams perfectly overlap, a composite beam with $l=+3$ is formed. Notably, there is an increase in the size of the vortex core compared to the $l=2$ case. This increase is attributed to the strong coupling between the size of the central dark core of conventional vortex beams such as LG and BG beams and their $l$. Hence, an increase in $l$ leads to a larger vortex core size. In contrast, POV beams exhibit an $l$-independent core size and radial intensity distribution . Throughout our numerical simulations, a net $l=+3$ is observed at all stages. Increasing the axial separation between the beams further results in the formation of two separate LG beams, each with $l=+3$.
The locations of the vortex cores of the composite LG beam can be found analytically by solving the equations $\mathrm{Re}[F_3(x,y)]=0$ and $\mathrm{Im}[F_3(x,y)]=0$ i.e.,

\begin{eqnarray}
    ((x-a)^{3}-3y^{2}(x-a))e^{\frac{2xa}{\omega_{0}^{2}}}+((x+a)^{3}-3y^{2}(x+a))e^{-\frac{2xa}{\omega_{0}^{2}}}=0\label{xC15}
\end{eqnarray}

and 

\begin{eqnarray}
    (3y(x-a)^{2}-y^{3})e^{\frac{2xa}{\omega_{0}^{2}}}+(3y(x+a)^{2}-y^{3})e^{\frac{-2xa}{\omega_{0}^{2}}}=0\label{yC16}.
\end{eqnarray}

At $x=0$, and for a large axial separation i.e., $a>0$, the roots of the Eq.~\ref{yC16} are evaluated as follows:

\begin{eqnarray}
    3ya^{2}-y^{3}+3ya^{2}-y^{3}&=&0\label{xC17}\\
    y(3a^{2}-y^{2})&=&0\label{yC18}
\end{eqnarray}
Therefore, from Eq.~\ref{yC18}, we found that $y=0$, and $y=\pm\sqrt{3}a$. As depicted in Fig.~\ref{F8}(a) for $\beta=0$, a vortex core of TC, $l=+3$, is located at $x=0$, and $y=0$. When the value of $\beta$ is set to 0.6, three vortices of TC, $l=+1$ , each are formed at: (1) $x=0$, $y=0$ , (2) $x=0$, $y=+\sqrt{3}a$, and (3) $x=0$,$y=-\sqrt{3}a$ (see Visualization 7). In the neighbourhood of the points $(0,\pm\sqrt{3}a)$, for smaller increments of $x$ and $y$, Eq.~\ref{xC14} can be approximated as
\begin{eqnarray}
    F_{3}(x=0+\delta x,y=\pm\sqrt{3}a+\delta y)\propto
    (3\omega_{0}^{2}-8a^{2})\delta x+3i\omega_{0}^{2}\delta y\label{yc19}.
    \end{eqnarray}

Thus, from Eq.~\ref{yc19}, it can be obtained that at $a=\sqrt{\frac{3}{8}}\omega_{0}$, $F_{3}(\delta y)\propto3i\omega_{0}^{2}\delta y$. This, in turn, implies that the intensity of the composite beam is negative (not shown here). As a result, the TC of the vortices located at $x=0,y=+\sqrt{3}a$, and $x=0,y=-\sqrt{3}a$ change from $l=+1$ to $l=-1$. Also, the cores located at $y=\pm\sqrt{3}a$ gets separated into two more vortices with TC, $l=+1$ at the origin. As a result of which, there are seven vortices in total (five with $l=+1$ each and two with $l=-1$ each) at $\beta=0.64$. This makes the net TC of the composite beam +3. For $a>\sqrt{\frac{3}{8}}\omega_{0}$, four vortices with $l=+1$ are always present in the composite beam (excluding the one at $x=0,y=0$). When the inter-axial separation increases further, i.e., for large values of $\beta$ (or, $a$), two distinct LG beams with TC, $l=+3$, are formed (see Fig.~\ref{F8}(c)). However, three vortices of TC , $l=-1$, are also present in the region where the field amplitude (or intensity) is nearly zero. As a result, the net TC of the composite beam is again +3. The numerically calculated line plots to locate the vortex cores are shown in Figs.~\ref{F8}(d)-(i) for three different values of $\beta$. The results obtained from the analytical approach can easily be verified with the line intensity plots. As shown in Figs.~\ref{F8}(d,g), a vortex core of TC, $l=+3$, is located at $x=0,y=0$ when $\beta=0$. As we change the value of $\beta$ further, vortex dynamics changes as evident from Figs.~\ref{F8}(e,h) and Figs.~\ref{F8}(f,i). At $\beta=2.0$, the composite LG beam decomposes into two well-separated LG beams with TC, $l+3$ each. There also exists three vortices with TC, $l=-1$ each, which makes the overall TC $+3$ at this stage (see Figs.~\ref{F8}(f,i)).

A comparison between the interference of high-order POV, and LG beams is discussed, and it is found that: (1) distinct interference patterns are formed in both cases, and (2) the vortex dynamics changes significantly, at same inter-axial separation distances. It is also observed that the splitting of the composite beam into its constituents is more rapid in the case of LG beams as compared to the POV beams. This feature can be attributed to the unique spatial profiles, these two beams possess. It is also interesting to note that both the analytical, and numerical methods can easily be applied to study the vortex dynamics in case of the composite LG beam, whereas, the numerical method is extremely useful for the description of the composite POV beam. The results obtained from both analytical, and numerical methods are consistent.

\section{Conclusions and Outlook}
In conclusion, our investigation delved into the interference of two parallel high-order POV beams, revealing that the positions of the vortex cores are contingent upon several factors: the axial separation between the beams, their relative phase shift, beam radius, and TC. Notably, the composite beam maintains an unchanged net TC, irrespective of the axial separation. However, at a larger axial separation, the composite beam divides into two distinct POV beams. A relative phase shift of $\Phi=\pi$ yields an interference pattern distinct from that of $\Phi=0$, yet the net TC content of the composite POV beam remains constant irrespective of different inter-axial separations. Increasing the TC did not yield significant differences in the interference pattern. An investigation into the effect of reducing the POV beam radius revealed two outcomes: a drastic change in the interference pattern and a more rapid splitting of the composite beam at a smaller axial separation. Comparing the interference of high-order POV and LG beams, it was found that their interference nature is similar, albeit with LG beams exhibiting a more rapid splitting of the composite beam into its individual components. We attribute this difference to variances in the spatial structure of the beams, despite both carrying OAM, and different vortex dynamics occurring in both these composite beams. Our findings are anticipated to be valuable in various technological applications, such as fiber-optic communication, where achieving high channel capacity and spectral efficiency is crucial. Specifically, it can aid in encoding useful information in different POV modes, coupled with optical fibers for transmission over long distances. It is also important to highlight that the use of multiplexed POV beams for optical communication minimizes the requirement of selective optics with specific parameters for vortices with different topological charge during both transmission, and reception in a communication network. Our theoretical analysis is limited to a certain position, i.e., at the waist, where both the beams are located. In future, it would indeed be interesting to explore the interference of high order OAM beams for changing propagation distances.

\begin{backmatter}
\bmsection{Funding} 
The present work is supported by the National Key Research and Development Program of China (Grant No.~2023YFA1407100), Guangdong Province Science and Technology Major Project (Future functional materials under extreme conditions - 2021B0301030005) and the Guangdong Natural Science Foundation (General Program project No. 2023A1515010871).

\bmsection{Disclosures} The authors declare no conflicts of interest.

\bmsection{Data Availability Statement} 
Data underlying the results presented in this paper are
not publicly available at this time but may be obtained from the authors upon reasonable request.

\end{backmatter}

%\section{References}

%Note that \emph{Optics Letters} and \emph{Optica} short articles use an abbreviated reference style. Citations to journal articles should omit the article title and final page number; this abbreviated reference style is produced automatically when the \emph{Optics Letters} journal option is selected in the template, if you are using a .bib file for your references.

%\bigskip
%\noindent Add citations manually or use BibTeX. See \cite{Zhang:14,OPTICA,FORSTER2007,testthesis,manga_rao_single_2007}. List up to three author names in references, and if there are more than three authors use \emph{et al.} after that.

% Bibliography
\bibliography{ApplOpt}

% Full bibliography added automatically for Optics Letters submissions; the following line will simply be ignored if submitting to other journals.
% Note that this extra page will not count against page length
\bibliographyfullrefs{ApplOpt}

%Manual citation list
%\begin{thebibliography}{1}
%\bibitem{Zhang:14}
%Y.~Zhang, S.~Qiao, L.~Sun, Q.~W. Shi, W.~Huang, %L.~Li, and Z.~Yang,
 % \enquote{Photoinduced active terahertz metamaterials with nanostructured
  %vanadium dioxide film deposited by sol-gel method,} Opt. Express \textbf{22},
  %11070--11078 (2014).
%\end{thebibliography}

% Please include bios and photos of all authors for aop articles
%\ifthenelse{\equal{\journalref}{aop}}{%
%\section*{Author Biographies}
%\begingroup
%\setlength\intextsep{0pt}
%\begin{minipage}[t][6.3cm][t]{1.0\textwidth} % Adjust height [6.3cm] as required for separation of bio photos.
%  \begin{wrapfigure}{L}{0.25\textwidth}
%    \includegraphics[width=0.25\textwidth]{john_smith.eps}
%  \end{wrapfigure}
%  \noindent
%  {\bfseries John Smith} received his BSc (Mathematics) in 2000 from The University of Maryland. His research interests include lasers and optics.
%\end{minipage}
%\begin{minipage}{1.0\textwidth}
%  \begin{wrapfigure}{L}{0.25\textwidth}
%    \includegraphics[width=0.25\textwidth]{alice_smith.eps}
%  \end{wrapfigure}
%  \noindent
%  {\bfseries Alice Smith} also received her BSc (Mathematics) in 2000 from The University of Maryland. Her research interests also include lasers and optics.
%\end{minipage}
%\endgroup
%}{}

\end{document}